\documentclass[english,aps,amssymb,prb,twocolumn,showpacs]{revtex4}
\pdfoutput=1
\usepackage[T1]{fontenc}
\usepackage[latin9]{inputenc}
\usepackage{amsmath}
\usepackage{graphicx}
\usepackage{amssymb}
\usepackage{esint}

\makeatletter
\@ifundefined{textcolor}{}
{% \definecolor{BLACK}{gray}{0}
 \definecolor{WHITE}{gray}{1}
 \definecolor{RED}{rgb}{1,0,0}
 \definecolor{GREEN}{rgb}{0,1,0}
 \definecolor{BLUE}{rgb}{0,0,1}
 \definecolor{CYAN}{cmyk}{1,0,0,0}
 \definecolor{MAGENTA}{cmyk}{0,1,0,0}
 \definecolor{YELLOW}{cmyk}{0,0,1,0}
 }

\renewcommand{\phi}{\varphi}
\renewcommand{\epsilon}{\varepsilon}

\renewcommand{\vec}[1]{{\bf #1}}
\newcommand{Û}{$}
\usepackage{epsfig}%\usepackage{showlabels}
\@ifundefined{showcaptionsetup}{}{%
\PassOptionsToPackage{caption=false}{subfig}}
\usepackage{subfig}
\makeatother

\usepackage{babel}

\begin{document}

\title {Helical Fermi arcs and surface states in time-reversal invariant Weyl semimetals}

\author{Teemu Ojanen}
\email[Correspondence to ]{teemuo@boojum.hut.fi}
%\author{Takuya Kitagawa$^2$}
\affiliation{Low Temperature Laboratory (OVLL), Aalto University, P.~O.~Box 15100,
FI-00076 AALTO, Finland }
%\affiliation{$^2$Physics Department, Harvard University, Cambridge, Massachusetts 02138, USA}
\date{\today}
\begin{abstract}
Weyl semimetals are gapless three-dimensional topological materials where two bands touch at even number of points in the Brillouin zone. In this work we study a zincblende lattice model realizing a time-reversal invariant Weyl semimetal. The bulk dynamics is decribed by twelve helical Weyl nodes.  Surface states form a peculiar quasi two-dimensional helical metal fundamentally different from the Dirac form typical for topological insulators. Allowed direction of velocity and spin of low-energy surface excitations are  locked to the cubic symmetry axes. The studied system illustrates general properties of surface states in systems with common crystal symmetries.

\end{abstract}
\maketitle
\bigskip{}

\section{Introduction}
Study of gapped topological matter, such as quantum Hall states, topological insulators (TI) and topological superconductors, has become a central topic in condensed matter physics.  \cite{kane2} Topological materials exhibit a wide range of exotic phenomena such as fractionalized excitations, quantized responses and unusual surface states that have become signature properties of these systems. A rapidly growing trend in search of new topological materials is the rising interest in gapless phases.  

Weyl semimetals are three-dimensional (3d) materials where two energy bands touch at even number of points in the reciprocal space.\cite{volovik} Band touching points, called Weyl nodes, are possible in systems where either inversion symmetry (IS) or time-reversal symmetry (TRS) is broken. Robustness of Weyl semimetals follows from topological properties of Weyl nodes acting as sources of the Berry curvature and defining monopoles in momentum space.\cite{volovik, volovik2} Characteristic physical properties of Weyl semimetals include unusual electromagnetic response\cite{hosur2, zyuzin1,zyuzin2, aji, vazifeh, liu, son, yang}  such as anisotropic DC conductivity and anomalous Hall effect and existence of Fermi arcs and topological surface states\cite{wan} and exotic broken symmetry states.\cite{cho2, Wei} There exists a number of promising theoretical proposals\cite{wan, witc, xu, burkov, burkov2, cho, yang, jiang, delplace} but no experimental verification has been achieved yet. While Weyl semimetals with broken TRS have been studied extensively, the time-reversal invariant case has received much less attention.\cite{halas} Weyl semimetals with unbroken TRS are not merely of academic interest since quite generally there exists a Weyl semimetal phase between a TI and a trivial insulator phases when IS is broken.\cite{murakami, yang2}            

In this work we study a time-reversal invariant tight-binding model in the zincblende lattice with a nearest neighbor hopping and a spin-orbit coupled next-nearest neighbor hopping. The orbitals on the different face-centered cubic (FCC) sublattices have different onsite energies, thus leading to IS breaking. The studied model is a variant of the Fu-Kane-Mele model realizing 3d TI phases.\cite{fu}  It is interesting that breaking IS by onsite potential leads to rich physics fundamentally different from the gapped TI phases.  

The studied model realizes a time-reversal invariant Weyl semimetal with twelve inequivalent Weyl nodes. In contrast to most of the previous studies where Weyl semimetals are realized by stacked 2d layers, the surface spectrum is identical for surfaces perpendicular to the three cubic axes. Populated surface states form a number of helical time-reversed patches bounded by Fermi arcs that are aligned with the cubic axes. Low-energy excitations in the vicinity of the Fermi arcs are highly unusual consisting of particles moving only along the cubic symmetry axes with their spin pointing to the perpendicular direction. Thus the surface spectrum provides a new type of helical metal very different from typical TI surface states \cite{kane2} and magnetic or irradiated \cite{ojanen} Rashba systems. Considering that much of the interest in time-reversal invariant TIs is concentrated on their helical surface metal, helical surface states on Weyl metals offer intriguing alternative realization with distinct properties. The studied model illustrates general features of Fermi arc structures of time-reversal invariant Weyl semimetal with common point-group symmetries.

\section{The studied model}
We are studying a variant of the tight-binding model on the zincblende lattice introduced by Fu, Kane and Mele in their pioneering work of topological insulators in three dimensions.\cite{fu}  The zincblende lattice consists of two FCC lattices displaced along the space diagonal by one quarter of its length. The Hamiltonian of the model is
\begin{align} \label{tb}
H=&-\sum_{\langle i, j \rangle} ( t\,c_i^{\dagger}c_j+\mathrm{h.c})+\sum_{ i} E_i\,c_i^{\dagger}c_i\nonumber\\
&+ i\lambda\sum_{\langle\langle i, j \rangle\rangle} ( c_i^{\dagger}\mathbf{e}_{ij}\cdot \mathbf{s} \,c_j-\mathrm{h.c}).
\end{align}
The first term corresponds to the nearest-neighbour hopping connecting points on different sublattices, the second term represents the staggered onsite potential and the last term is the spin-orbit hopping between the second neighbours residing on the same sublattice characterized by $\lambda$. Following Ref.~ \onlinecite{fu}, we have chosen a next-nearest  spin-orbit hopping term  which respects IS in addition to TRS.  We assume that the lattice sites are distinguished by their onsite potential so that $E_i=\pm\epsilon$ depending on which sublattice the site resides. This breaks IS of the diamond structure and is crucial for the appearance of the Weyl semimetal phase. The spin-orbit field is determined by the two bond vectors connecting the second neighbours as $\mathbf{e}_{ij}=\frac{\mathbf{e}_{i}\times\mathbf{e}_{j}}{|\mathbf{e}_{i}\times\mathbf{e}_{j}|}$ and $\mathbf{s}=(s_x,s_y,s_z)$ are the Pauli matrices acting on spin.  The staggered onsite potential breaks IS which is crucial in obtaining the Weyl semimetal phase. The model (\ref{tb}) gives rise to the Bloch Hamiltonian
\begin{align} \label{bloch}
H=d_1(k)\sigma_x+d_2(k)\sigma_y+\epsilon\sigma_z+\nonumber\\
\sigma_z\left[D_x(k)s_x+D_y(k)s_y+D_z(k)s_z\right],
\end{align}
with 
\begin{align} \label{d}
&d_1(k)=t(1+\cos{\vec{k}\cdot\vec{a}_1}+\cos{\vec{k}\cdot\vec{a}_2}+\cos{\vec{k}\cdot\vec{a}_3)}\nonumber\\
&d_2(k)=t(\sin{\vec{k}\cdot\vec{a}_1}+\sin{\vec{k}\cdot\vec{a}_2}+\sin{\vec{k}\cdot\vec{a}_3})\nonumber\\
&D_x(k)=\lambda\left[(\sin{\vec{k}\cdot\vec{a}_2}-\sin{\vec{k}\cdot\vec{a}_3}-\right.\nonumber\\ 
&\left.\sin{(\vec{k}\cdot\vec{a}_2 -\vec{k}\cdot\vec{a}_1)}+\sin{(\vec{k}\cdot\vec{a}_3 -\vec{k}\cdot\vec{a}_1)}\right])
%&D_y(k)=\lambda\left[(\sin{\vec{k}\cdot\vec{a}_2}-\sin{\vec{k}\cdot\vec{a}_3}-\sin{(\vec{k}\cdot\vec{a}_2 -\vec{k}\cdot\vec{a}_1)}+\sin{(\vec{k}\cdot\vec{a}_3 -\vec{k}\cdot\vec{a}_1)}\right])\nonumber\\
%&D_z(k)=\lambda\left[(\sin{\vec{k}\cdot\vec{a}_2}-\sin{\vec{k}\cdot\vec{a}_3}-\sin{(\vec{k}\cdot\vec{a}_2 -\vec{k}\cdot\vec{a}_1)}+\sin{(\vec{k}\cdot\vec{a}_3 -\vec{k}\cdot\vec{a}_1)}\right])
\end{align}
and $(\sigma_x,\sigma_y,\sigma_z)$ are the Pauli matrices acting on sublattice space. 
The primitive and reciprocal vectors of the FCC lattice are given by
\begin{align} \label{lat1}
\vec{a}_1&=\frac{a}{2}\,(0,1,1)\qquad\vec{b}_1=\frac{2\pi}{a}\,(-1,1,1)\nonumber\\
\vec{a}_2&=\frac{a}{2}\,(1,0,1)\qquad\vec{b}_2=\frac{2\pi}{a}\,(1,-1,1)\nonumber\\
\vec{a}_3&=\frac{a}{2}\,(1,1,0)\qquad\vec{b}_3=\frac{2\pi}{a}\,(1,1,-1).
\end{align}
The other components $D_{y/z}(k)$ are obtained by permuting  lattice vectors cyclically in the expression for $D_x(k)$.  The time-reversal operation acting on Bloch Hamiltonians and states has a representation $T=-is_yK$ where $K$ denotes complex conjugation. The studied model (\ref{bloch}) satisfies $TH(k)T^{-1}=H(-k)$, which is the criteria for a model to be time-reversal invariant.    

\section{The Weyl node structure} 
\begin{figure}
%\centering
\includegraphics[width=0.6\columnwidth]{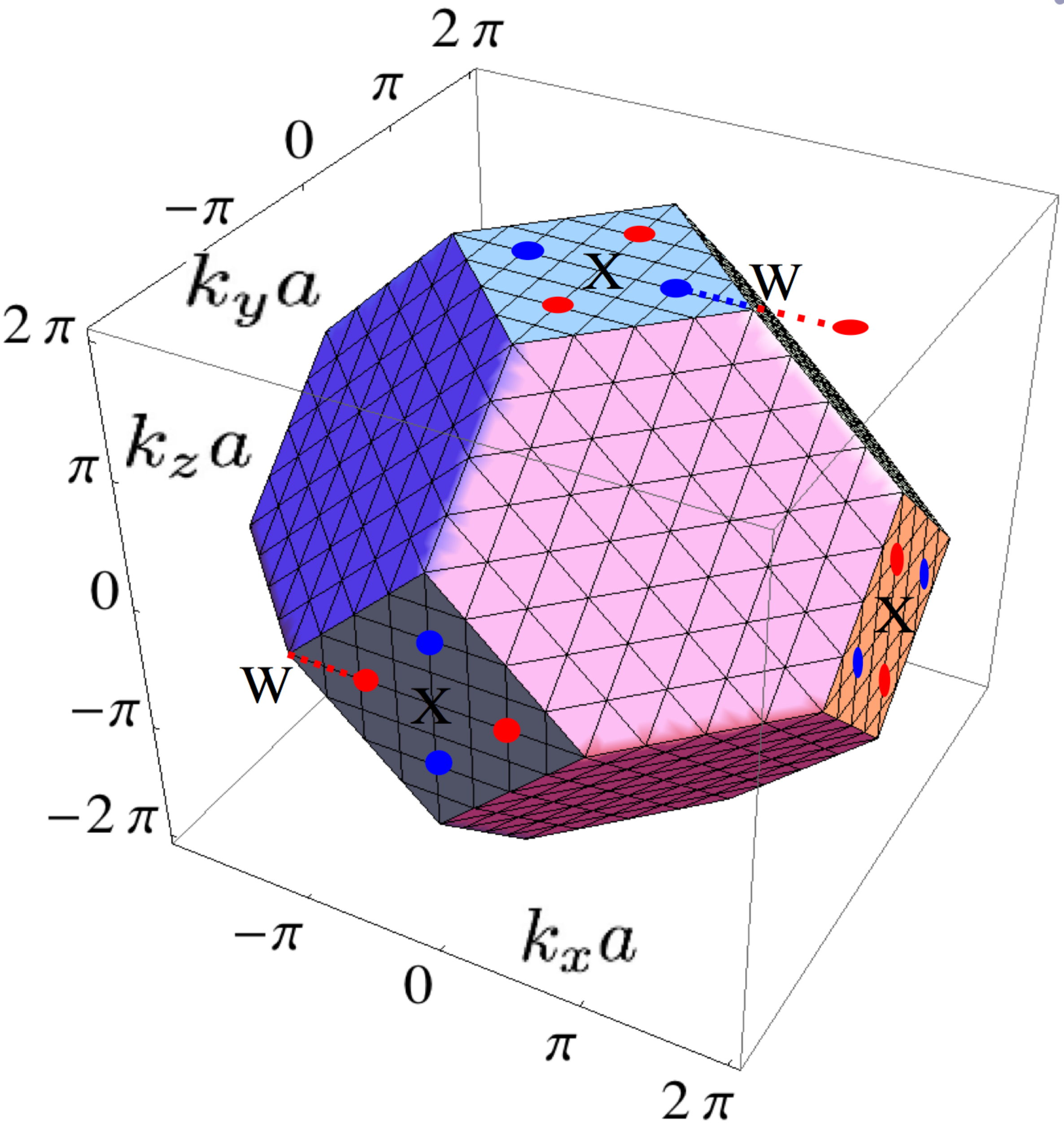}
\caption{ Red and blue circles represents the twelve independent Weyl nodes of opposite chirality in the first Brillouin zone.  The node outside the zone is equivalent to the one in the $k_z=0$ plane and illustrates how the nodes annihilate at critical value $\epsilon=\epsilon_c$.}\label{fig1}
\label{sceme}
\end{figure}

The effective Weyl Hamiltonian can be found by reducing the four band model to a gapless and gapped two band models. This can be achieved by identifying the $k$-dependent eigenstates of  the spin sector $\frac{ \mathbf{D}\cdot \mathbf{s}}{D}|\pm D\rangle=\pm|\pm D \rangle$ where  $ \mathbf{D}=(D_x(k),D_y(k),D_z(k))$ and $D=\sqrt{\sum_iD_i(k)^2}$. In this basis the spin part can be replaced by its eigenvalues in (\ref{bloch}), leading to two band Hamiltonians for these spin sectors,   
 \begin{align} \label{weyl1}
H_{\pm}=d_1(k)\sigma_x+d_2(k)\sigma_y+(\pm D(k)+\epsilon)\sigma_z.
\end{align}
Assuming $\epsilon>0$ ($\epsilon<0$),  $H_+$ ($H_-$) is always gapped but band touching in the $H_-$ ($H_+$) sector may occur if  $d_1=d_2=0$ and $D=\epsilon$ $(D=-\epsilon)$ for some $k=k_0$. When the IS breaking potential is weaker than the critical value $\epsilon<\epsilon_c=4|\lambda|$ the band-touching conditions yield twelve inequivalent Weyl nodes in the first Brillouin zone at $(\pm k_0,0,\frac{2\pi}{a})$,  $(0,\pm k_0,\frac{2\pi}{a})$,  $(\pm k_0,\frac{2\pi}{a},0)$,  $(0,\frac{2\pi}{a},\pm k_0)$,$(\frac{2\pi}{a},0,\pm k_0,)$,  $(\frac{2\pi}{a},\pm k_0,0)$, where $\sin{\frac{k_0}{2}}=\frac{\epsilon}{4\lambda}$. The Weyl node structure is illustrated in Fig.~\ref{fig1}.  At $\epsilon=0$ the system is gapless and the Weyl nodes of opposite helicity sit on top of each other. When inversion symmetry is weakly broken $0<|\frac{\epsilon}{4\lambda}|\ll 1$, the nodes move away from the three inequivalent X points $(0,0,\frac{2\pi}{a})$, $(0,\frac{2\pi}{a},0)$ and $(\frac{2\pi}{a},0,0)$. By increasing $|\frac{\epsilon}{4\lambda}|$, the Weyl nodes move toward the four inequivalent $W$ points on the zone boundary. Since the $W$ points on different faces are connected by reciprocal lattice vectors $\vec{b}_i$ they annihilate each other at the critical value $|\frac{\epsilon}{4\lambda}|=1$, beyond which the system becomes a trivial insulator. The dashed lines illustrate this process for a pair of nodes. One way to understand the topological properties of the system is to map the Hamiltonians (\ref{weyl1}) to the Bloch sphere. In the gapped phase when $|\epsilon|>\epsilon_c$ it is not possible to cover the whole sphere since $(\pm D(k)+\epsilon)$ has always the same sign when $k$ takes all values in the Brillouin zone. Particularly, if $k$ is restricted to a 2d submanifold defining a 2d Brillouin zone, the Chern number of the mapping is always zero for both spin sectors. The insulator phase is trivial without protected surface states.

\begin{figure}
%\centering
\includegraphics[width=0.6\columnwidth]{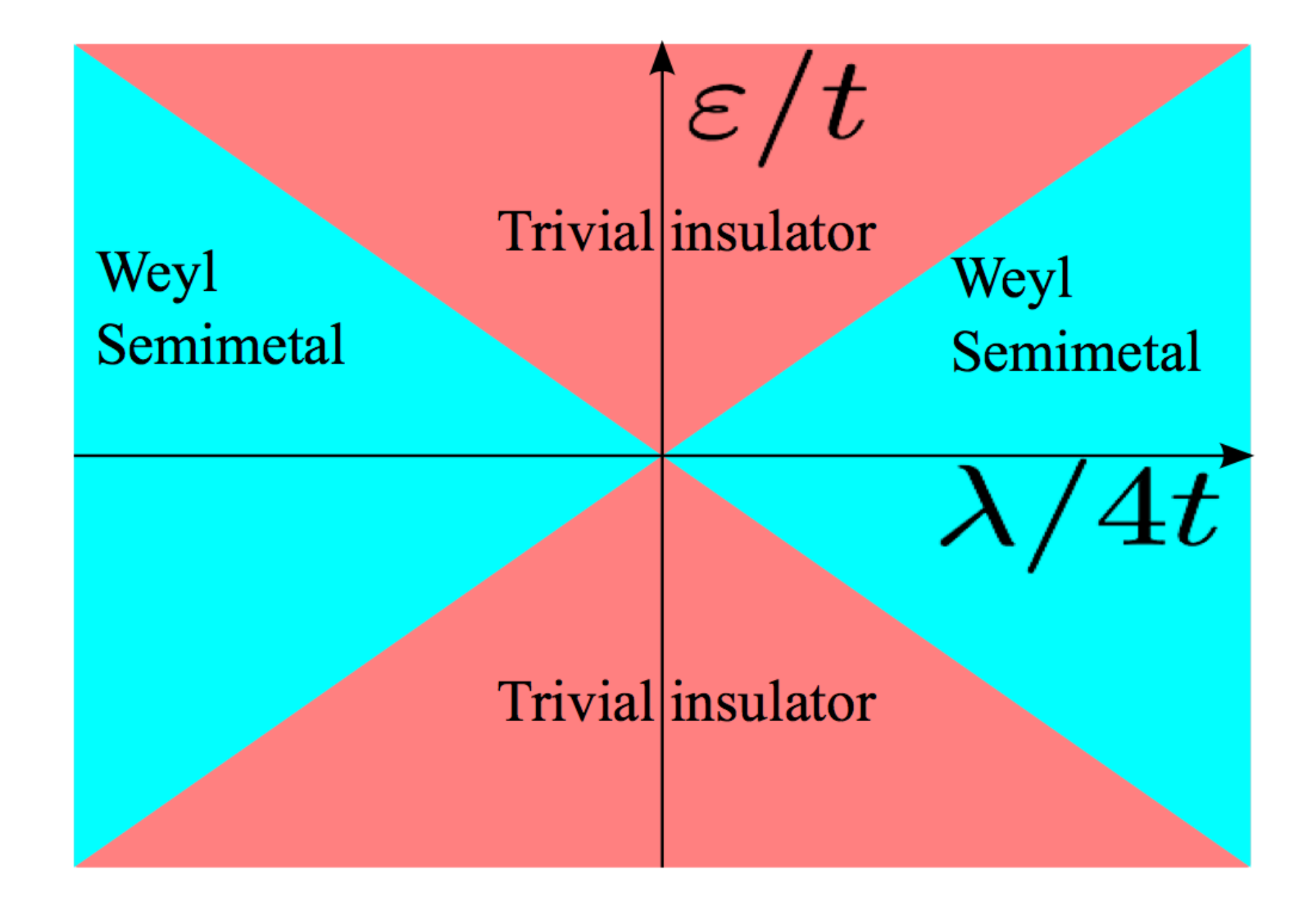}
\caption{ Phase diagram as a function of the IS-breaking sublattice potential $\epsilon$ and spin-orbit coupling $\lambda$. At $\epsilon=0$ line inversion symmetry is restored and the system is a Dirac semimetal with a four-fold degenerate touching of two conduction and two valence bands at $X$ points. }\label{figx}
\label{sceme}
\end{figure}
On general grounds it is known that six of the twelve nonequivalent Weyl nodes have positive (negative) helicity and that nodes related by TRS have the same helicity. A two-fold rotational symmetry along cubic axes imply that the nodes connected by $\pi$ rotation have the same helicity. The zincblende structure also respects $\frac{\pi}{2}$ rotations combined with a reflection with respect to the plane orthogonal to the rotation axis. The nodes related by this symmetry have opposite helicities due to reflection. These expectations can be verified in a systematic approach. In the vicinity of a Weyl node the effective Hamiltonian can be brought to the form $H=v_{x}k_x\sigma_x+v_{y}k_y\sigma_y+v_{z}k_z\sigma_z$ describing a hedgehog in the reciprocal space.\cite{volovik} The topological invariant associated with a Weyl point can be calculated as a flux through the surface containing the hedgehog and takes values $\nu=\mathrm{sign} \left(v_{x}v_{y}v_{z}\right)$. For a general Hamiltonian $H=\sum_in_i(k)\sigma_i$ with a Weyl node at $k=k_0$ the invariant can be evaluated by  $ \nu=\mathrm{sign}\left[\mathrm{det}\left(\partial_{k_i}n_j(k)\right)\right]_{|k=k_0}$. Employing this formula we can now identify the helicities of Weyl points yielding $\nu_{(\pm k_0,0,\frac{2\pi}{a})}=-\mathrm{sign}\,{\epsilon}$, $\nu_{(0,\pm k_0,\frac{2\pi}{a})}=\mathrm{sign}\,{\epsilon}$, $\nu_{(\frac{2\pi}{a},\pm k_0,0)}=-\mathrm{sign}\,{\epsilon}$, $\nu_{(\frac{2\pi}{a},0,\pm k_0)}=\mathrm{sign}\,{\epsilon}$, $\nu_{(\pm k_0,\frac{2\pi}{a},0)}=\mathrm{sign}\,{\epsilon}$, $\nu_{(0,\frac{2\pi}{a},\pm k_0,)}=-\mathrm{sign}\,{\epsilon}$. As illustrated in Fig.~\ref{fig1}, this helicity structure confirms expectations from symmetry considerations. The gapless sector of Eq.~(\ref{weyl1}) describes a 3d helical semimetal with opposite spin textures near nodes connected by TRS. Generally, if Weyl nodes are grouped into pairs of opposite helicities, the sum of the resulting separation
vectors vanishes (modulo a resiprocal lattice vector) by virtue of TRS. 

The existence of a Weyl semimetal phase in the studied model can be understood by arguments  due to Murakami.\cite{murakami} At a generic (non-TRS) point in the Brillouin zone the bands are non-degenerate, so a generic crossing happens between two bands. The codimension of this crossing is three since the most general Hamiltonian in the two-band subspace is described by a linear combination of the three Pauli matrices (and an unimportant term proportional to the unit matrix).  Because the dimensionality of the system coincides with the codimension of the crossing, the touching of the bands may take place in isolated points in the Brillouin zone. There exist a finite region where a small variation of one parameter in the effective $2\times 2$ Hamiltonian,  such as  $\epsilon$ in the studied case, just shifts the band touching points but cannot gap them. Large perturbations may gap the system by bringing bringing two nodes of opposite helicity together and annihilating them. General phase diagram of a system with broken IS  is more complicated,\cite{yang2} but it is reasonable to expects a finite gapless region between gapped phases in the studied model as illustrated in Fig.~\ref{figx}.

\section{surface states and Fermi arcs}

In the presence of a surface, Weyl semimetals exhibit  topologically protected surface states.  Projections of Weyl nodes to the surface determine end points of Fermi arcs, a line in the surface Brillouin zone separating occupied and empty states. The Fermi arc structure determines low-energy properties of surface excitations. It is expected that Fermi arcs are formed between the nodes that annihilate each other at $W$ points since a trivial insulator does not support surface states. 

Assuming the IS breaking parameter satisfies $\epsilon>0$, the Weyl nodes are contained in the $H_-$ sector with spin wavefunction $|- D\rangle$. 
First we study the surface spectrum when the system is terminated by a $(001)$ plane. The in-plane momenta $k_x,k_y$ remain good quantum numbers and parametrize the spectrum in the surface Brillouin zone $|k_x\pm k_y|\leq \frac{2\pi}{a}$. The Fermi arcs in this direction follow from the horizontally moving Weyl nodes, the arcs from the vertically moving nodes shrink to points in the surface Brillouin zone boundary.  As illustrated in Fig.~\ref{fig1}, the horizontally moving nodes have lattice equivalent counterparts in the  $k_z=\frac{2\pi}{a}$ plane so we can concentrate on the neighborhood of $H_-(k_z=\frac{2\pi}{a})$.  In order to solve the surface modes, we define a new variable $q=(k_x,k_y,\frac{2\pi}{a})$ and linearize $H_-$ in $\tilde{q}_z\equiv k_z-\frac{2\pi}{a}$ around $\tilde{q}_z=0$ which yields   
\begin{align} \label{lin2}
H(q)&=d_1(q)\sigma_x+d_2(q)\sigma_y+\left[ -D(q)+\epsilon\right]\sigma_z+\nonumber\\
&+\tilde{q}_z \left[d_1'(q)\sigma_x+d_2'(q)\sigma_y\right] +\mathcal{O}[\tilde{q}_z^2],
\end{align}
where $d_1'(k)=\partial_{k_z}d_1(k_x,k_y,k_z)_{|k_z=\frac{2\pi}{a}}$ and  $d_2'(k)=\partial_{k_z}d_2(k_x,k_y,\frac{2\pi}{a})$.
Defining $d'(q)\equiv\sqrt{d_1'^2+d_2'^2}$ and $d(q)\equiv\sqrt{d_1^2+d_2^2}$ and employing the orthogonality $d_1'(q)d_1(q)+d_2'(q)d_2(q)=0$, it follows that a rotation in the sublattice space
$U=e^{-i\theta\sigma_z/2}$ with $\sin\,\theta=\frac{d_2'}{d'}$ transforms Eq.~(\ref{lin2}) to
\begin{align} \label{lin3}
H=\tilde{q}_zd'(q)\sigma_x+d(q)\mathrm{sign}(k_xk_y)\sigma_y+\left[ \epsilon-D(q)\right]\sigma_z.
\end{align}
The sign function determines whether $(d_1',d_2',0)$, $(d_1,d_2,0)$ and $(0,0,1)$ defines a right or left-handed orthogonal set.
 
By substituting $\tilde{q}_z\to -i\partial_z$ in Eq.~(\ref{lin3}), we search for normalizable surface states satisfying $H\Psi=E(k_x,k_y)\Psi$. Before proceeding further we must determine the appropriate boundary condition that describes the vacuum-Weyl semimetal boundary. The system is gapped when $\epsilon>\epsilon_c$ and as discussed above, this phase corresponds to a topologically trivial band insulator. Choosing coordinates so that $z=0$ coincides with the boundary, the surface states can found by requiring that  $\epsilon=\epsilon_+>\epsilon_c$ for $ z>0$  (outside the system) and $\epsilon=\epsilon_-<\epsilon_c$ for $z<0$ (inside the system). From the topological point of view vacuum can be regarded as a band insulator with a very large band gap, so the above boundary condition is appropriate to describe a Weyl metal terminated by a surface.\cite{burkov}

\begin{figure}
%\centering
\includegraphics[width=0.45\columnwidth, clip=true]{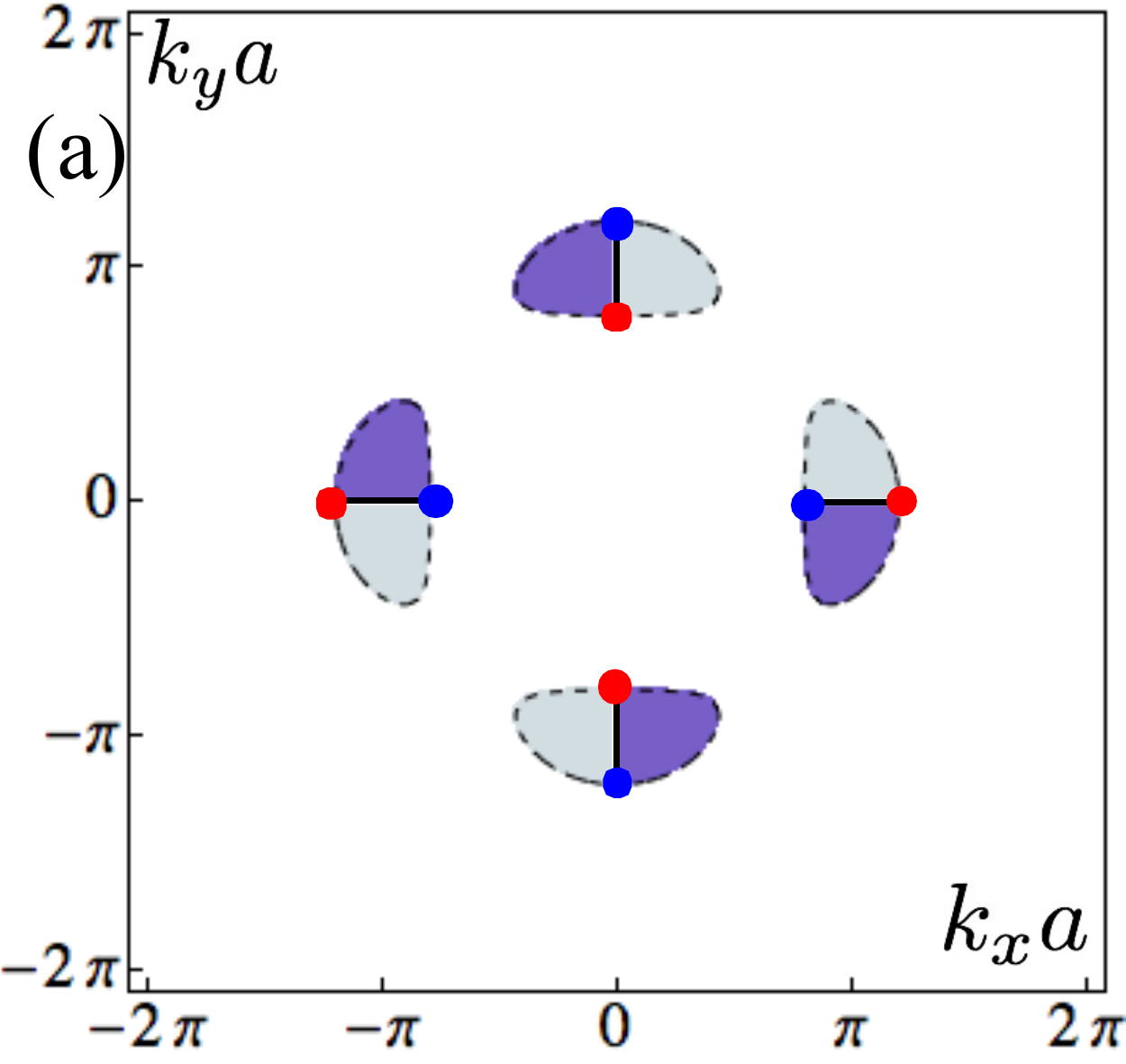}
\includegraphics[width=0.45\columnwidth, clip=true]{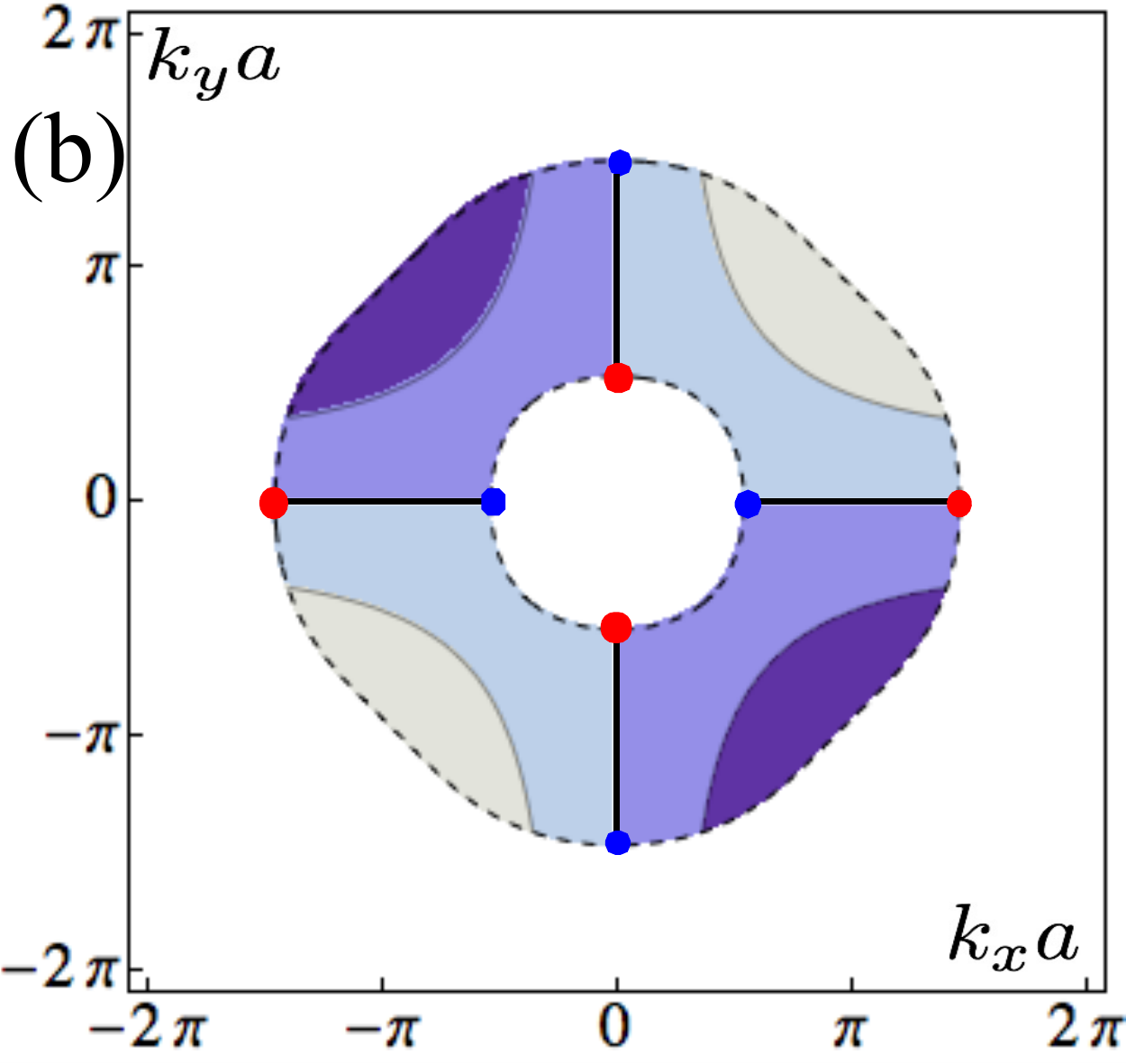}
\includegraphics[width=0.45\columnwidth, clip=true]{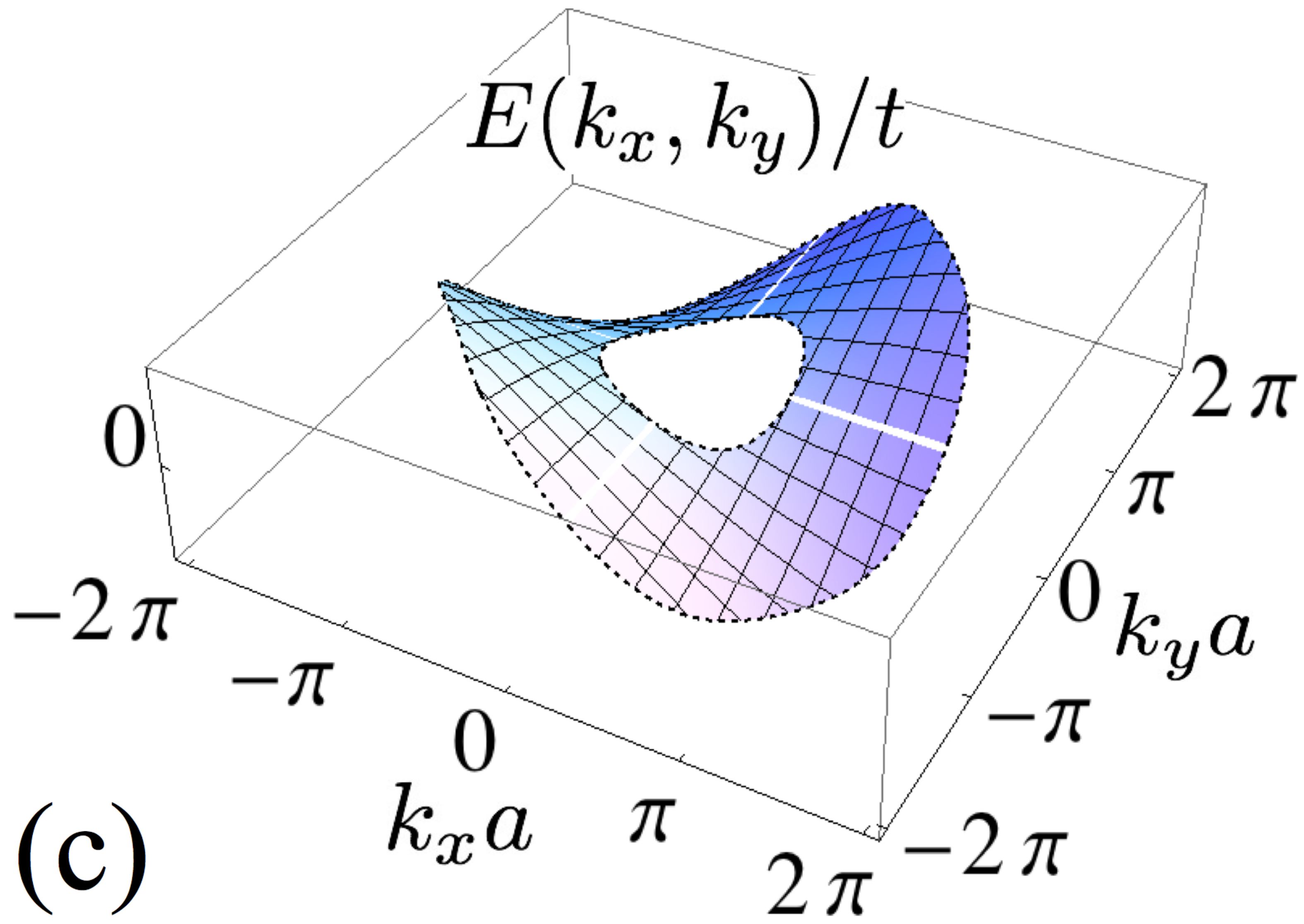}
\includegraphics[width=0.45\columnwidth, clip=true]{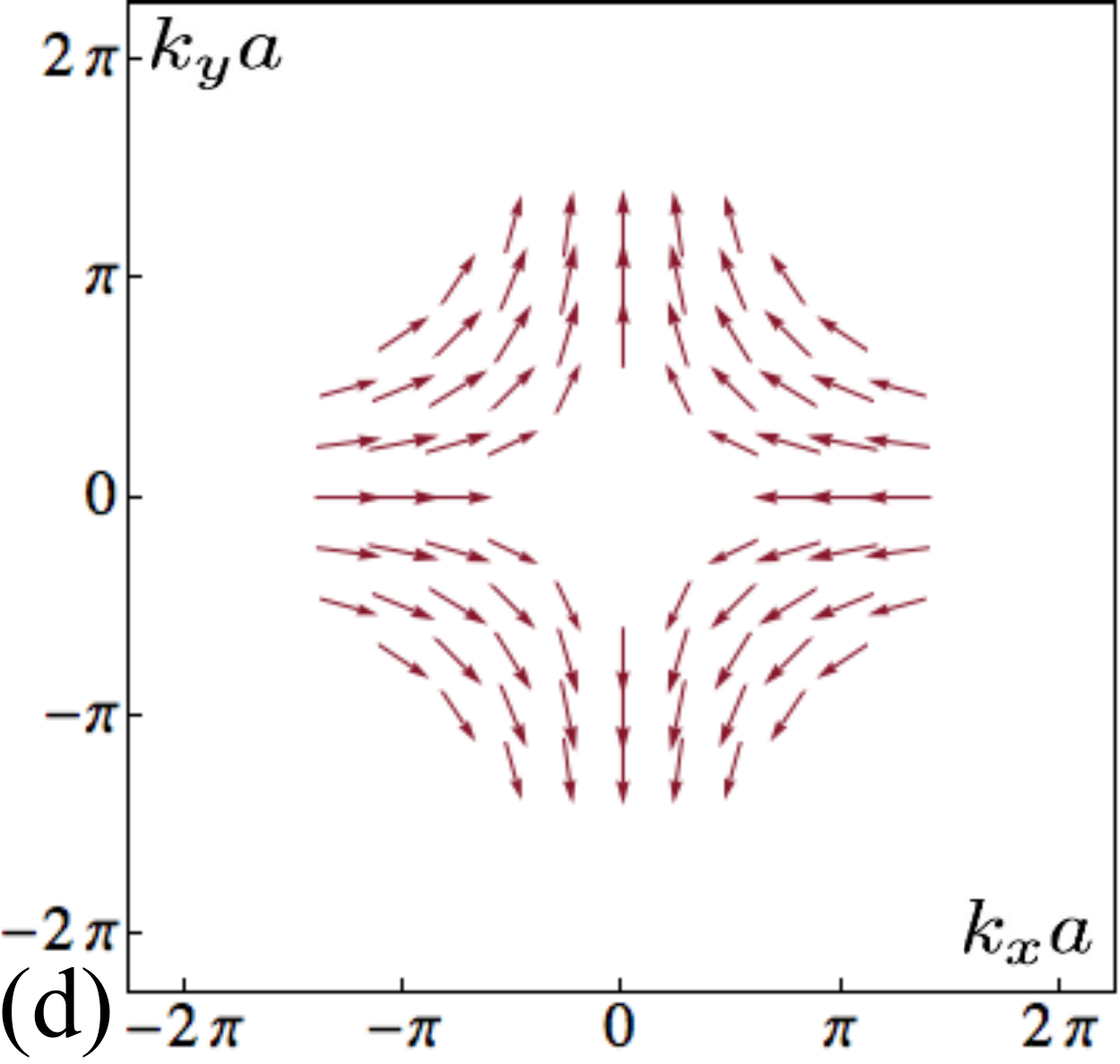}
\caption{(a): The surface states first appear on separate small patches when the system is close to the band insulator transition $\frac{\epsilon}{4\lambda}=0.95$  and grow together when the system enter deeper into the Weyl metal phase. The lines connecting the nodes represent the Fermi arcs. On the the white area the penetration depth perpendicular to the surface diverges. (b): Same as (a) but for $\frac{\epsilon}{4\lambda}=0.75$. (c): Energy dispersion of the surface states for $\frac{\epsilon}{4\lambda}=0.75$. (d): Spin texture of the surface states. On the Fermi arcs spin points in one of the four allowed directions.}
\label{fig2}
\end{figure}

For positive $\epsilon$ we find a state localized on the surface for each  $(k_x, k_y)$ satisfying $D(q)>\epsilon_-$ given by
\begin{align} \label{sspec3}
\Psi(k_x,k_y,z)&=Ae^{\mp\frac{z}{\xi_{\pm}}} |+ y\rangle|- D\rangle,
\end{align}
where upper (lower) signs correspond to region outside (inside) the system.
Here $|+y\rangle$ denotes the state for which $\sigma_y|+y\rangle=|+y\rangle$ and we have included the spin part of the wavefunction explicitly. The momentum-dependent penetration depth of wavefunctions into the bulk is given by $\xi_{\pm}=\pm\frac{d'(q)}{-D(q)+\epsilon_\pm}$ (which is positive under our assumptions). The dispersion of the surface states is given by
\begin{align} \label{sspec1}
E(k_x,k_y)=\mathrm{sign}(k_xk_y)d(q)=4t\sin{\frac{k_xa}{4}}\sin{\frac{k_ya}{4}}.
\end{align}
As plotted in Figs.~\ref{fig2} (a) and (b), the region for allowed surface states are bounded by curves where the penetration depth $\xi_-$ diverges (the black dashed curves). The divergence of $\xi_-$ is accompanied by the merging of the surface spectrum to the bulk spectrum so the surface states cannot be distinguished from the bulk states anymore.  The positive (negative) energy states exist in disjoint, time-reversed patches of the surface Brillouin zone. More precisely, the states at $k$ and $-k$ have the same energy and opposite spin. The Fermi arcs connecting the Weyl points and satisfying condition $E(k_x,k_y)=0$ coincide with the coordinate axes $k_x=0$ and $k_y=0$ separating the positive and negative energy surface patches as illustrated in Figs.~\ref{fig2} (a), (b) and (c).  Spin of the surface states, shown in Fig.~\ref{fig2} (d), is parallel to $(-D_x(k), -D_y(k), 0)_{|k_z=\frac{2\pi}{a}}$ and have only an in-plane component. The surface states are helical: the in-plane momentum determines the spin state uniquely. However as discussed below, surface states are fundamentally different from the Dirac form encountered in many TIs.

Allowing also negative values of $\epsilon$, the surface states exists for $(k_x, k_y)$ satisfying $D(k_x, k_y,\frac{2\pi}{a} )\geq |\epsilon_-|$. The spectrum and eigenfunctions are given by those presented above with the exception that energy changes its sign and spin and sublattice states are flipped $|- D\rangle\to |+ D\rangle $, $|+y\rangle\to |-y\rangle $. Therefore the surface spectrum is not continuous at $\epsilon=0$. This behaviour is similar to the discontinuity in the anomalous Hall insulator described by a Dirac equation with sign-changing mass term. Although the spectrum is not continuous at zero mass, the penetration depth of the surface states diverge and smoothens out the transition in finite systems.      
The solution for the surface spectrum by employing the linearized (long wavelength) Hamiltonian (\ref{lin3}) is valid as long as the characteristic spatial variation is smooth $\xi_-\gg a$. This condition fails at the surface Brillouin zone boundary since $d'(q)=0$  at $k_y=k_x\pm\frac{2\pi}{a}$. However, for large enough $\epsilon$ values the surface spectrum does not extend to the Brillouin zone boundary and analytical approximation provides accurate description everywhere. Also, for weakly broken IS $\frac{\epsilon}{\lambda}\lesssim (\frac{a}{\xi_-})^2$ the approach fails since in that case the two spin sectors (\ref{weyl1}) become nearly degenerate and the long-wavelength approximation cannot resolve which one contains the Weyl nodes. However, the case of small  $\epsilon$ is of limited interest since the surface states are not well-localized to surface anyway.

The above analysis identifies Fermi arcs and surface states for $(001)$ surfaces.  A similar analysis as presented above can be carried out for $(010)$ and $(100)$ surfaces and as the symmetric structure of the nodes in Fig.~\ref{fig1} suggests, the spectrum for those surfaces are essentially identical to the spectrum of the $(001)$ surface. 

\section{discussion and summary}
As discussed above, the low energy sector of surface states consist of excitations in the vicinity the two pairs of time-reversed Fermi arcs. By time-reversed Fermi arcs we mean arcs connected by transformation $k\to-k$. Since the electronic dispersion is flat along the Fermi arcs,  velocity $v= \partial_kE(k)$ is perpendicular to the arc and spin near the arcs as illustrated in Figs.~\ref{fig2} (c) and (d). Interestingly, velocity and spin can have only one of the four allowed orientations near the arcs. Therefore the allowed low-energy surface excitations resemble to those of two perpendicular 1d helical metals rather than a 2d helical metal realized on the surfaces of many TI materials. Time-reversed Fermi arcs also posses a perfect nesting property generic for 1D systems. The surface state structure could be probed by spin and angular resolved photoemission spectrum techniques or by electron tunnelling spectroscopy as discussed in Ref.~\onlinecite{hosur1}.  

One naturally wonders which properties of the studied model are generic and carry to other Weyl semimetals with unbroken TRS. The key properties determining the Fermi arc structure and spin textures on Fermi arcs are TRS and crystal symmetries. Let us assume that the system is terminated by a plane which admits a perpendicular axis of n fold rotations and m fold rotations combined with a reflection. From TRS we know that if there exists a Fermi arc between nodes at $k_1$ and $k_2$ there must exist another arc terminated by nodes at $-k_1$ and $-k_2$ with identical helicities, but opposite spin. Since the system is invariant in n fold rotations, there must also exist n time-reversed Fermi arc pairs. Spin on the arc is also rotated but helicities of rotated nodes remain the same. Improper rotations also lead to m time-reversed Fermi arc pairs but the nodes connected by this transformation have opposite helicities. The spin state remains unaffected in reflection.  The Fermi arc and spin structure in Figs.~\ref{fig2} (b) and (c) are in agreement with the two-fold rotational symmetry and four-fold rotation combined with reflection. The arcs obtained by two-fold rotations and time-reversal coincide, so there is only two pairs of arcs.  It is now quite straightforward to imagine allowed Fermi arc structures in systems with different point-group symmetries.  
%\section{Summary}

In summary, we studied a lattice model of a time-reversal invariant Weyl semimetal and its helical surface Fermi arcs. The surface states are identical in all three surfaces perpendicular to the cubic axes of the zincblende lattice. The low-energy excitations at Fermi arcs consist of helical electrons with discrete allowed velocity and spin orientations determined by the cubic axes. The surface states can be regarded as a new type of helical metal fundamentally different from the Dirac form realized in typical topological insulator surfaces.     

It is pleasure to thank Grigori Volovik and Pauli Virtanen for many illuminating discussions. The author acknowledges the Academy of Finland for support.

\end{document}